\documentclass[]{spie}  

\usepackage{amsmath,amsfonts,amssymb}
\usepackage{graphicx}
\usepackage[colorlinks=true, allcolors=blue]{hyperref}

\usepackage{mathtools}

\usepackage[colorinlistoftodos]{todonotes}

\title{Overcoming the low signal-to-noise problem for hybrid mode-selective photonic lantern-based wavefront correction using machine learning}

\author[a]{Nathan~K.~Long}
\author[a,b]{Barnaby~Norris}
\author[a]{Daniel~S.~Dahl}
\author[b]{Christopher~H.~Betters}
\author[a]{Julia~J.~Bryant}
\author[c]{Nick~Cvetojevic}
\author[b]{Sergio~Leon-Saval}
\author[c]{Frantz~Martinache}
\author[c]{Marc-Antoine~Martinod}
\author[a]{Akira~Rodziewicz-Ryan}
\author[d]{Adam~K.~Taras}
\author[b]{Jin~Wei}
\author[a,b]{Peter~G.~Tuthill}
\affil[a]{Astralis-USyd, Sydney Institute for Astronomy (SIfA), School of Physics, University of Sydney, Camperdown, NSW, Australia}
\affil[b]{Sydney Astrophotonics Instrumentation Laboratories, Sydney Institute for Astronomy (SIfA), University of Sydney, Camperdown, NSW, Australia}
\affil[c]{Observatoire de la Cote d’Azur, Université Côte d’Azur, Nice, France}
\affil[d]{Leiden University, Leiden, Netherlands}

\authorinfo{N.K.L. e-mail: nathan.long@sydney.edu.au}

\begin{document} 
\maketitle

\begin{abstract}
Hybrid mode-selective photonic lanterns transform an input complex point-spread function into several single-mode outputs, where a selected core feeds the fundamental mode to a photonic science instrument, while the remaining cores are used for wavefront sensing in a closed-loop adaptive optics system. A neural network maps the intensities of the wavefront sensing cores to an estimated wavefront correction, which is applied to an upstream deformable mirror. However, there exists a trade between maximizing the amount of light reserved for the photonic instrument and the reduced signal-to-noise ratios for the wavefront sensing cores. We explore wavefront correction for the Seidr instrument, a part of the Asgard Suite for the Very Large Telescope Interferometer. We evaluate different neural network architectures, comparing wavefront estimation performance for different wavefront error types, as a first step toward addressing the signal-to-noise trade-off. Results show transformer neural networks as a promising solution for temporal photonic lantern-based wavefront estimation.
\end{abstract}


\section{INTRODUCTION}
\label{sec:intro}  

Photonic lanterns (PLs) are photonic devices designed to transform an electric field from a multi-mode fiber (MMF) input to several single-mode fiber (SMF) outputs, where the electric field is decomposed into spatial `lantern' modes~\cite{Leon2013}. The output core intensities can then be used for wavefront sensing in adaptive optic (AO) systems~\cite{Norris2020}.

Conventional wavefront sensors require light from the pupil plane to be separated into a path for the scientific instrument and a path to image a beam\textquoteright s wavefront, such that non-common path aberrations (NCPAs) arise outside of the sensor\textquoteright s ability to detect. Hybrid mode-selective PLs (HMS-PLs) overcome this limitation by guiding the fundamental LP01 mode into a mode-selective core. The light in the mode-selective core can then be injected into a subsequent science instrument. Simultaneously, the light in the rest of the cores can be used for wavefront sensing, where the wavefront sensing is in the same final focal plane as the science instruments, such that the NCPAs can be measured and removed~\cite{Norris2022_hybrid}.

PL wavefront correction algorithms must learn the transformation between the phase wavefront at the pupil plane and the intensities measured from the HMS-PLs output cores. Machine learning offers a solution, such that the PL output intensities are used as input to a neural network (NN), which then estimates a wavefront correction to be applied to an upstream deformable mirror (DM) (e.g., ~\cite{Norris2020}).

However, given that the objective of HMS-PLs is to maximize light in the mode-selective core, light is thereby minimized in the wavefront sensing cores, resulting in lower signal-to-noise ratios (SNRs). As a first step towards addressing this problem, we present an analysis of two different neural network architectures designed for wavefront estimation using simulated data for a HMS-PL.

We present our analysis in the context of the Seidr instrument~\cite{Taras2024_seidr, Long2026_seidr}. Seidr is a kernel-nulling interferometer being developed as a node of the Asgard Instrumentation Suite~\cite{Martinod2023, Taras2024_heimdallr, Courtney2024_baldr, Garreau2024_nott}, at the Very Large Telescope Interferometer (VLTI). Seidr provides a technological demonstration of kernel-nulling, and of the use of HMS-PLs for high contrast and high angular resolution imaging of hot young Jovians, circumstellar dust, and provides a path towards exomoon detection~\cite{Taras2024_seidr}.

\textcolor{black}{We compare the wavefront estimation performance of a transformer NN (TNN) with the performance of a convolutional NN (CNN), for several different wavefront error types, providing the first analysis of TNNs for PL-based wavefront estimation.} First, we compare the wavefront estimations for tip/tilt wavefront aberrations, which are assumed to dominate the beams pre-Seidr, with Von K\'arm\'an seeing estimation, assuming no upstream AO correction. Second, we compare wavefront estimation performance of the two networks for randomized instances of wavefront errors with correlated temporal sequences of wavefront errors.

We provide a description of our simulation framework in Section~\ref{sec:sims}, we define the NN-based wavefront estimation algorithms in Section~\ref{sec:tnn}, we analyze the results in Section~\ref{sec:perf}, and give concluding remarks in Section~\ref{sec:concl}.

\section{SIMULATION SETUP}
\label{sec:sims}

\textcolor{black}{HMS-PLs are designed to guide the fundamental LP01 mode from a point-spread function (PSF) at the MMF input through a \textit{mode-selective} core, optimizing light injection into a downstream science instrument~\cite{Norris2022_hybrid}. The light in the higher-order modes is guided to the rest of the \textit{wavefront sensing} cores, which can be used to estimate wavefront corrections to apply to an upstream DM.}

\begin{figure}[ht] 
    \begin{center}
    \includegraphics[scale=0.9]{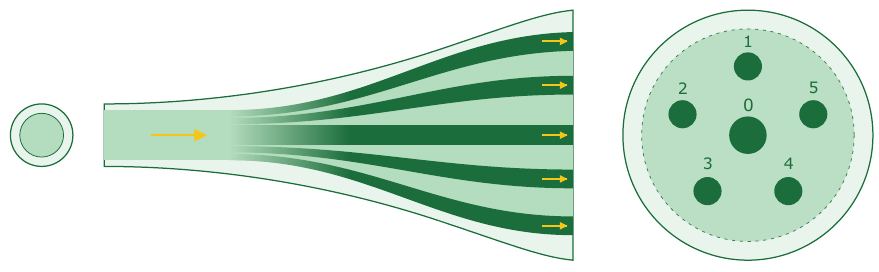}
    \end{center}
    \caption{Schematic of a 6-core hybrid mode-selective photonic lantern.}
    \label{fig:pl}
\end{figure}

\textcolor{black}{A schematic of the 6-core HMS-PL simulated in this work is given in Figure~\ref{fig:pl}. The mode-selective core is the central core~0, while the wavefront sensing cores are arranged in a surrounding pentagonal configuration (cores~1~to~5). The shade of green represents the different refractive indices in the lantern, where darker green represents a higher refractive index, and the yellow arrows show the direction of light propagation.}

To understand the potential performance of HMS-PLs at wavefront sensing to maximize light injection into the mode-selective core, we simulate a stellar source-to-lantern output, where our wavefront estimation algorithms are trained and tested using the resulting data.

A point source, with a wavelength of 1550~nm, is propagated through Von K\'arm\'an seeing (representing atmospheric distortion) or as post-AO corrected tip-tilt wavefront aberrations, described in Section~\ref{sec:wfe}. The distorted electric field is transmitted through the telescope aperture and propagated to the focal plane, where the resulting PSF at the lantern input is computed under the assumption of Fraunhofer diffraction. The PSF ${E_S(x,y)}$ is then decomposed into the guided LP mode electric fields ${E_{LP}(x,y)}$, where their complex coefficients $C_{lm}$ are calculated using the normalized inner product, 
\begin{equation} \label{eq:c_lm}
    C_{lm} = \frac{\iint E_S(x,y) \cdot E^*_{LP}(x,y) \ dxdy}{\left[(\iint ||E_S(x,y)||^2 \ dxdy) \cdot (\iint ||E_{LP}(x,y)||^2 \ dxdy)\right]^{1/2}}.
\end{equation}

\begin{figure}[ht] 
    \begin{center}
    \includegraphics[scale=0.6]{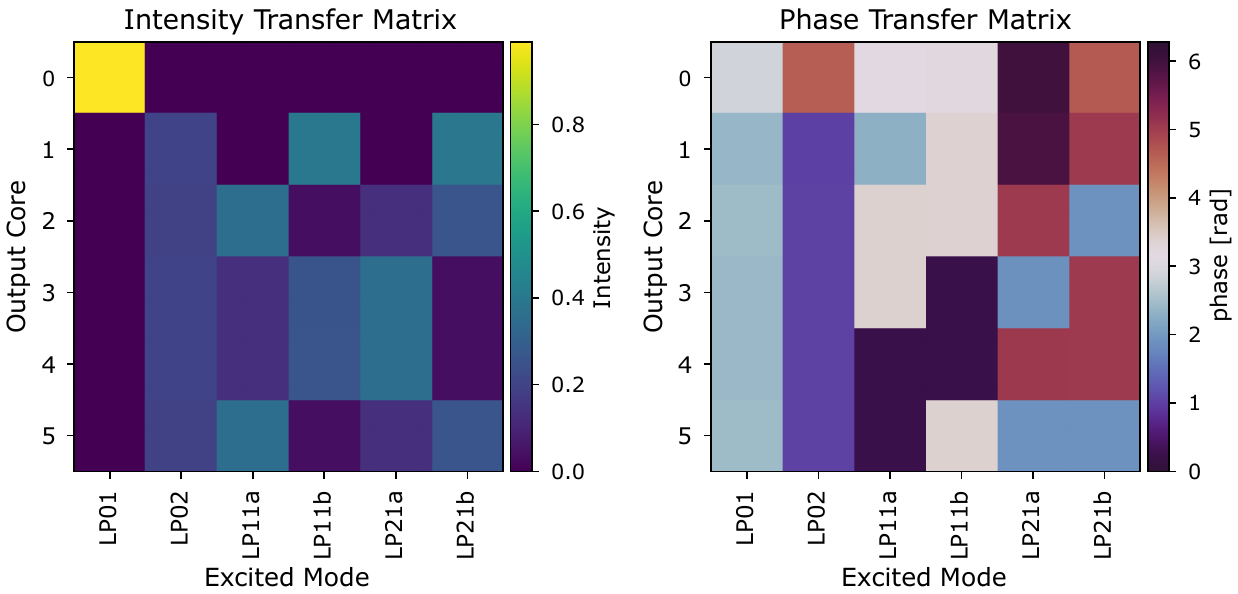}
    \end{center}
    \caption{6-core hybrid mode-selective photonic lantern intensity (left) and phase (right) transfer matrices for $\lambda=1550$~nm.}
    \label{fig:tf}
\end{figure}

\noindent The resulting complex coefficients are propagated through a 6-core HMS-PL using a pre-defined transfer matrix, as seen in Figure~\ref{fig:tf}. \textcolor{black}{The transfer matrix represents the intensity and phase transformation of the guided LP modes injected into the MMF end of the lantern to the output cores at the SMF end.} The transfer matrix used in this work is calculated using a finite-difference beam propagation method~\cite{Pedrola2015} with adaptive meshing~\cite{Shibayama1999}.

It can be seen that the LP01 mode is strongly guided in the central core~0, while the higher-order modes are guided in the wavefront sensing cores~1~to~5. The lantern properties used in this work are given in Table~\ref{tab:pl_param}. Of note, mode-selectivity is achieved by doubling the radius of the mode-selective core, with respect to the wavefront sensing cores, while the refractive indices are kept the same.

\begin{table}[ht]
    \caption{Hybrid mode-selective photonic lantern properties.}
    \label{tab:pl_param}
    \begin{center}
    \begin{tabular}{|l|c|}
        \hline
        \rule[-1ex]{0pt}{3.5ex} \textbf{Parameter} & \textbf{Value} \\ \hline
        \rule[-1ex]{0pt}{3.5ex} PL length & 50,000~$\mu$m \\ \hline
        \rule[-1ex]{0pt}{3.5ex} WFS SMF core radii & 4.1~$\mu$m \\ \hline
        \rule[-1ex]{0pt}{3.5ex} MS SMF core radius & 8.2~$\mu$m \\ \hline
        \rule[-1ex]{0pt}{3.5ex} Output cladding radius & 155~$\mu$m \\ \hline
        \rule[-1ex]{0pt}{3.5ex} Output core spacing & 102.5~$\mu$m \\ \hline
        \rule[-1ex]{0pt}{3.5ex} Taper ratio & 20 \\ \hline
        \rule[-1ex]{0pt}{3.5ex} WFS core refractive indices & 1.449 \\ \hline
        \rule[-1ex]{0pt}{3.5ex} MS core refractive indices & 1.449 \\ \hline
        \rule[-1ex]{0pt}{3.5ex} Cladding refractive index & 1.444 \\ \hline
        \rule[-1ex]{0pt}{3.5ex} Capillary refractive index & 1.435 \\ \hline
        \end{tabular}
    \end{center}
\end{table}

\subsection{Wavefront Errors} \label{sec:wfe}

We investigate two different conditions of wavefront errors to understand the potential performance of HMS-PLs for wavefront sensing and mode-selective light injection. 

Turbulent phase screens are generated using the Von K\'arm\'an spatial phase power spectral density (PSD) $\Xi_{atm}$, defined by,
\begin{equation} \label{eq:von_kar}
    \Xi_{atm}(\kappa) = 0.023 \ r_0^{-5/3} \frac{1}{(\kappa^2 + \kappa^2_0)^{11/6}},
\end{equation}

\noindent where $r_0$ is the Fried parameter, $\kappa$ is the spatial angular frequency, $\kappa_0 = 2\pi/L_0$, for the outer scale $L_0$~\cite{Schmidt2010}. Here, we set ${r_0 = 0.4}$~m and ${L_0 = 10}$~m. The resulting distorted complex electric field approximates a stellar source after propagation through the Earth\textquoteright s atmosphere before reaching a ground-based telescope (here, the VLTI). The PSFs generated using Von K\'arm\'an seeing assume that there is no upstream adaptive optics correction before light is injected into the HMS-PL. Temporal sequences of phase screens are generated using Taylor\textquoteright s frozen turbulence hypothesis~\cite{Taylor1938}, with a transverse wind speed of~10~m/s. \textcolor{black}{It is worth noting that the simulated 6-core HMS-PL only guides six~LP modes, thereby providing only a low-order wavefront sensing capability.}


\textcolor{black}{In the case of Seidr, there are two upstream adaptive optics systems --- NAOMI (for the ATs) or GPAO (for the UTs), followed by Baldr (within the Asgard Suite). However, given the likely vibration sensitivity of the beam paths within BIFROST, in this work, we assume that the wavefront errors encountered by the HMS-PLs in Seidr are dominated by tip-tilt aberrations.}

We generate the tip/tilt wavefront errors using the 2nd and 3rd Zernike polynomials. Zernike polynomials $Z_n^m(\rho, \beta)$ are an infinite set of polynomials, which can be used to represent any phase profile~\cite{Noll1976}, and are defined for a unit circle, with polar coordinates $\rho$ and $\beta$ as,
\begin{equation} 
    Z_n^m(\rho, \beta)=
    \begin{cases*}
        R_n^m(\rho) \cos(m\beta),       & \ \text{if $m\geq0$}, \\
        R_n^{-m}(\rho) \sin(-m\beta),   & \ \text{otherwise},
    \end{cases*}
\end{equation}

\noindent where ${n - m}$ must be even, for the integers $m$ and $n$, and ${n \geq m \geq 0}$. The radial polynomial term $R_n^m(\rho)$ is then given by,
\begin{equation}\label{eq:zernike_poly}
  R_n^m(\rho) = \sum^{\frac{n-m}{2}}_{j=0} \frac{(-1)^j (n-j)!}{j! \left(\frac{n+m}{2} -j\right)! \left(\frac{n-m}{2} -j\right)!} \rho^{n-2j}.
\end{equation}

\noindent A tip/tilt phase profile $\Delta\Phi(\rho, \beta)$ can be represented by the summation of the weighted Zernike polynomials, using the Zernike coefficients $a^m_n$ as,
\begin{equation}\label{eq:zernike_phase}
  \Delta\Phi(\rho, \beta) = a^1_1 Z^1_1(\rho, \beta) + a^{-1}_1 Z^{-1}_1(\rho, \beta),
\end{equation}

\noindent for the tip/tilt indices of $n=1$, $m=\pm1$. Note that $0 \leq \rho \leq 1$ and $0 \leq \beta \leq 2\pi$. 

Temporal sequences of phase screens are generated by smoothly varying the Zernike coefficients as follows. A vector $\hat{x}_t$ of sequential length $N_t$ is sampled from a unit Gaussian distribution $\mathcal{N}(0,1)$ and convolved with a Gaussian kernel to produce $\hat{y}_t$, where $t$ represents the time step. Each $\hat{y}_t$ is normalized by its standard deviation and scaled to an RMS error of~100~nm.

\section{WAVEFRONT ESTIMATION ALGORITHMS} \label{sec:tnn}

Current photonic lantern wavefront estimation algorithms are primarily based on multi-layer perceptrons (e.g., ~\cite{Norris2020, Sweeney2021}). In this work, we compare the estimation performance of two different NN types, CNNs and TNNs, where we aim to show the performance of each algorithm for both randomized and temporal wavefront errors, as well as both Von K\'arm\'an and tip/tilt wavefront error types. 

\subsection{Convolutional Neural Network}

CNNs are traditionally designed to estimate two-dimensional data, where relationships between array elements are exploited. We develop a decoder CNN using three main layer types: fully-connected layers, convolutional layers, and upsampling layers. Fully-connected layers form weighted relationships between perceptrons (nodes) in one layer and the perceptrons in the next layer. In the convolutional layers, a kernel is passed over sub-arrays within the data, extracting features between neighboring elements. Upsampling layers expand the spatial dimensions of the data using bilinear interpolation. A more detailed description of convolutional neural network operations is given in~\cite{Krizhevsky2012}.

\begin{figure}[h] 
    \centering
    \includegraphics[scale=0.8]{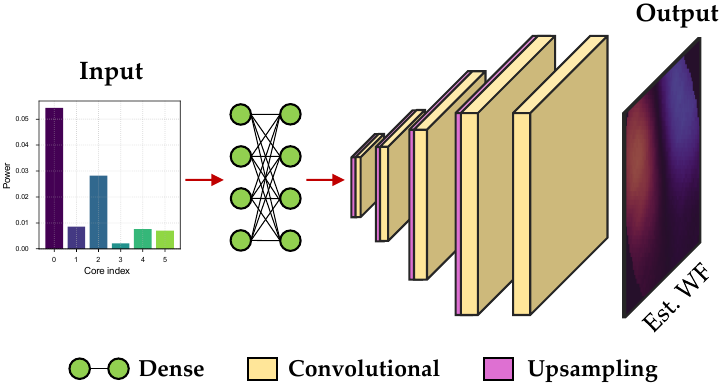}
    \caption{Convolutional neural network architecture designed for hybrid mode-selective photonic lantern wavefront estimation.}
    \label{fig:cnn}
\end{figure}

Figure~\ref{fig:cnn} depicts the CNN architecture designed for HMS-PL wavefront estimation. The HMS-PL core intensities are first passed through two fully-connected layers, resulting in a feature vector, which is reshaped into a low-resolution \textit{spatial feature array} for convolutional decoding. The feature array is then upsampled and convolved across four upsampling + convolutional blocks, progressively increasing the spatial resolution while extracting features between neighboring elements. The data from the final upsampling + convolutional block is passed through a final convolutional layer, before outputting a wavefront estimation. 

The CNN hyperparameters are given in Table~\ref{tab:cnn_params}. The dense expansion width $n_{dense}$ defines the number of units in the first fully-connected layer, while the base convolutional filter count $n_{filt}$ sets the number of filters in the first upsampling + convolutional block. The filter count is halved after each subsequent block, down to a minimum of eight filters. The number of upsampling + convolutional blocks $n_{ups}$ determines how many times the spatial resolution is doubled, expanding an initial $4\times4$ feature volume to the full $64\times64$ wavefront pixel resolution. A dropout rate $D$ of~0.2 is applied after the first fully-connected layer and after each upsampling + convolutional block for regularization.

\begin{table}[ht]
\caption{Convolutional neural network architecture hyperparameters.}
\label{tab:cnn_params}
\begin{center}
\begin{tabular}{|l|c|}
\hline
    \rule[-1ex]{0pt}{3.5ex} \textbf{Hyperparameter} & \textbf{Value} \\ \hline
    \rule[-1ex]{0pt}{3.5ex} Base convolutional filters $n_{filt}$ & 64 \\ \hline
    \rule[-1ex]{0pt}{3.5ex} Dense expansion units $n_{dense}$ & 128 \\ \hline
    \rule[-1ex]{0pt}{3.5ex} Upsampling steps $n_{ups}$ & 4 \\ \hline
    \rule[-1ex]{0pt}{3.5ex} Dropout $D$ & 0.20 \\ \hline
\end{tabular}
\end{center}
\end{table}

\subsection{Transformer Neural Network}

TNNs use \textit{self-attention} to learn relationships between elements in an input sequence. In this work, the input consists of temporal sequences of lantern core intensities. Positional encoding is added to the embedded sequence, distinguishing the ordering of the temporal intensities. The positionally encoded modes are then passed through the attention layers. Each attention layer contains multiple attention heads, which calculate \textit{query}, \textit{key}, and \textit{value} interpretations of the input data, calculating \textit{attention weights} which describe the similarity between the data across different time steps. Using multiple attention heads allows different temporal relationships within the lantern intensity sequence to be learned. The outputs of the attention heads are then passed through fully-connected layers, which further transform the learned information and map it to the desired wavefront estimation. A more detailed description of transformer neural networks is given in~\cite{Vaswani2017}.

\begin{figure}[ht]
    \begin{centering}
    \includegraphics[scale=0.765]{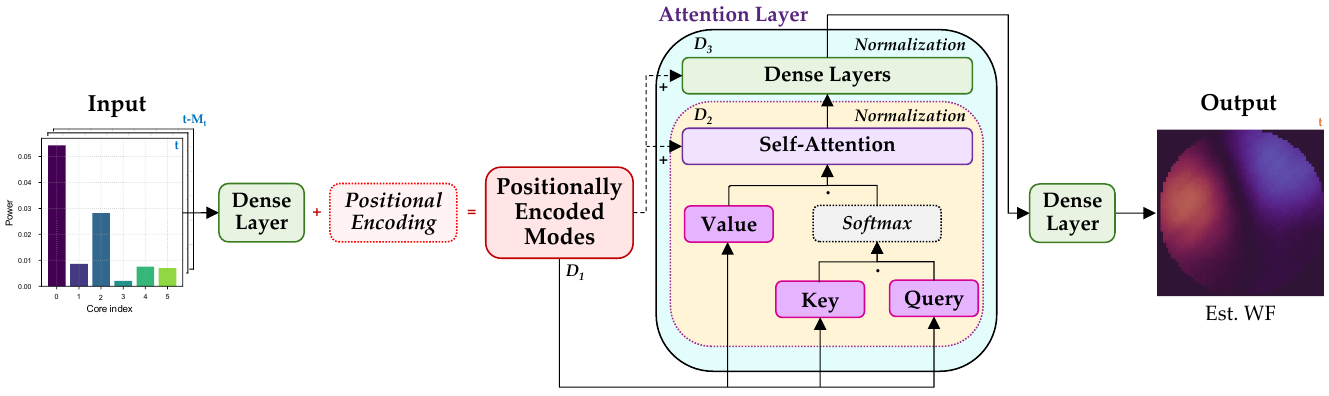} 
    \caption{TNN architecture designed for hybrid mode-selective photonic lantern wavefront estimation. Sequence of previous $M_t$ lantern intensities taken as input, then a wavefront estimate is output for the current time-step $t$. Transformer layer architecture is depicted in the light blue block, while the attention head architecture is depicted in the yellow block.}\label{fig:tnn}
    \end{centering}
\end{figure}

Figure~\ref{fig:tnn} depicts the TNN architecture designed for HMS-PL wavefront estimation. The input vector is first passed through a fully-connected (dense) layer, resulting in a higher-dimensional embedding of the PL mode outputs. Positional encoding is performed on the fully-connected layer outputs, resulting in \textit{positionally encoded modes}, which are input to the attention layers. The data output from the attention layers is then passed into a final fully-connected layer, before outputting a wavefront estimate.

The TNN hyperparameters are given in Table~\ref{tab:tnnseq_params}. The transformer width $d_{m}$ flows through the entire network, defining the initial and final fully-connected layers, the positional encoding, the key dimensions (${d_m//n_{heads}}$), and the second fully-connected layer within the attention layers. The intermediate dimension $d_{ff}$ defines the length of the first fully-connected layer within the attention layers. The dropout rate after the positionally encoded mode is set to 0.1, while the self attention layers, and fully-connected layers within the attention layers are set to~0.2.

\begin{table}[ht]
\caption{Transformer neural network architecture hyperparameters.}
\label{tab:tnnseq_params}
\begin{center}
\begin{tabular}{|l|c|}
\hline
    \rule[-1ex]{0pt}{3.5ex} \textbf{Hyperparameter} & \textbf{Value} \\ \hline
    \rule[-1ex]{0pt}{3.5ex} Transformer width $d_{m}$ & 128 \\ \hline
    \rule[-1ex]{0pt}{3.5ex} Intermediate dimension $d_{ff}$ & 256 \\ \hline
    \rule[-1ex]{0pt}{3.5ex} Attention heads $n_{heads}$ & 4 \\ \hline
    \rule[-1ex]{0pt}{3.5ex} Transformer layers $\#$ & 2 \\ \hline
    \rule[-1ex]{0pt}{3.5ex} Dropout (positional) $D_1$ & 0.10 \\ \hline
    \rule[-1ex]{0pt}{3.5ex} Dropout (attention) $D_2$ & 0.20 \\ \hline
    \rule[-1ex]{0pt}{3.5ex} Dropout (fully-connected) $D_3$ & 0.20 \\ \hline
\end{tabular}
\end{center}
\end{table}

Note that, in the case of randomized wavefront errors, the TNN architecture is modified to only input a single PL intensity vector (rather than a time-series).

\subsection{Training and Operation}

We simulate~100,000 instances of the point source - to - lantern intensity outputs for each wavefront error type, storing the lantern intensity and wavefront data. The 100,000 instances are then split into 70,000 training instances, 15,000 validation instances, and 15,000 testing instances.

During training, the known input lantern intensities are mapped to the known output wavefronts, such that the weights within the CNN and TNN are trained to learn the relationship between the lantern inputs and wavefront outputs. Both models are trained using the Adam optimizer~\cite{Kingma2015_adam} to minimize a mean squared error loss function, for~500 epochs (with an early stopping patience of 50~epochs) and a batch size of~64. After training, all weight values remain fixed.

During operation, the wavefront estimation algorithms take known (measured) lantern intensities as input, then output an estimate of the causal wavefront, which can be applied to an upstream DM. We analyze the open-loop case, without considering the effects of latency between the output estimation and difference in wavefront when applying the corrections (a current focus of our ongoing work for Seidr).

In the case of randomized phase screens, both the CNN and TNN take a single (`one shot') lantern intensity measurement as input to output an estimate of the wavefront correction. In the case of the temporal phase screens, the CNN takes a single (`one shot') lantern intensity measurement as input, while the TNN takes the previous ${M_t=50}$ lantern intensity measurements as input, to estimate the wavefront correction for the current time step.

Without compiling the NN models as TensorRTs, or running an optimization on their hyperparameters, we provide an example of the inference latency of each NN model. Inference latency represents the computational time required for the NNs to output a single wavefront estimate. Using a single NVIDIA~GeForce~RTX~4090~GPU, the CNN has a mean inference latency of~0.375~ms, while the non-sequential TNN has a latency of~0.649~ms and the sequential TNN has a latency of~0.685~ms.

\section{WAVEFRONT ESTIMATION PERFORMANCE} \label{sec:perf}

Understanding the wavefront estimation capability of the 6-core HMS-PLs for different wavefront error conditions permits a greater understanding of the potential improvements in mode-selective light injection (e.g., for increased light injection into the kernel-nulling chip in Seidr), informing future design decisions. 

We first analyze the tip/tilt wavefront estimation performance of the TNN and CNN, which offers a better representation of the likely dominant wavefront aberrations encountered by Seidr. We then analyze the \textcolor{black}{low-order} wavefront estimation performance of the TNN and CNN when encountering Von K\'arm\'an seeing, with no upstream AO correction. The phase screens for both wavefront error types are generated both as randomized independent instantiations and as temporal sequences (described in Section~\ref{sec:wfe}).

\subsection{Tip/Tilt Wavefront Aberrations}

We set the RMS error of the tip/tilt wavefront aberrations to 100~nm to analyze the low-order wavefront estimation performance of the CNN and TNN for both randomized and temporal data.

\begin{figure}[ht] 
    \centering
    \includegraphics[scale=0.55]{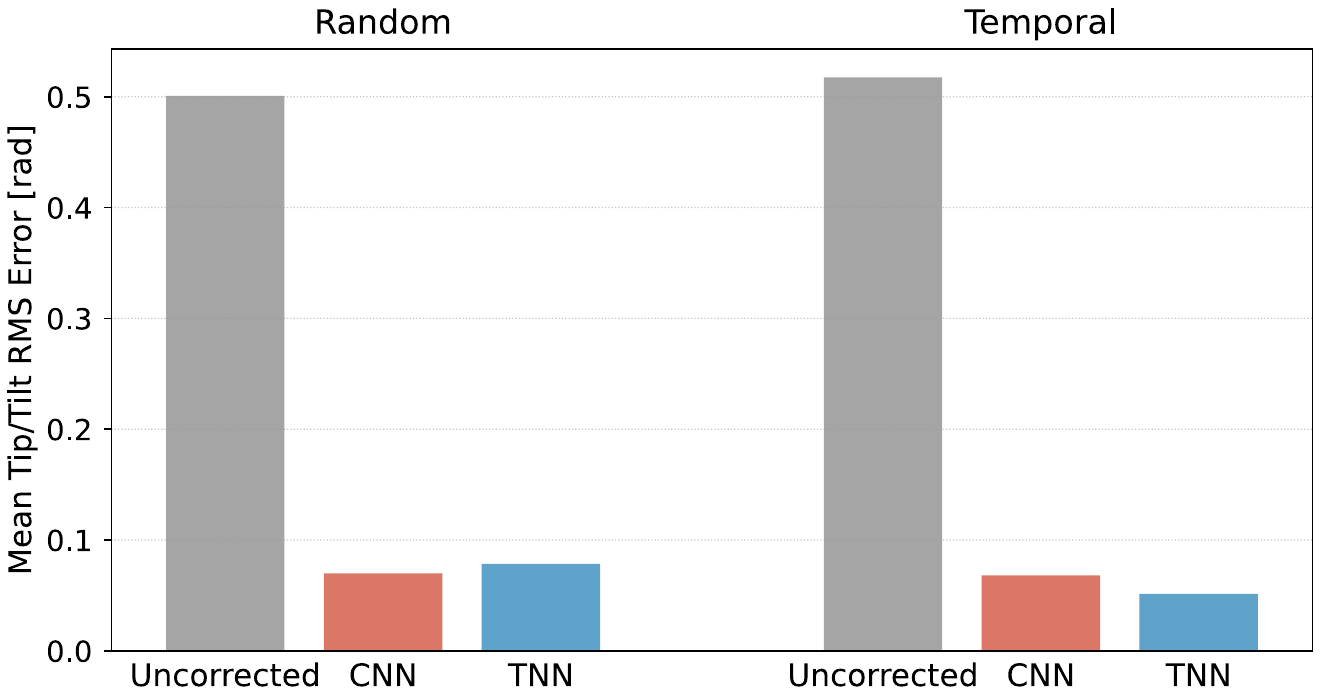}
    \caption{Tip/tilt wavefront RMS error before and after wavefront estimation applied to correct wavefront.}
    \label{fig:tt_rms}
\end{figure}

Figure~\ref{fig:tt_rms} shows the mean tip/tilt RMS error for the uncorrected wavefront (grey), and the mean tip/tilt RMS error for the CNN (red) and TNN (blue) when the estimated corrections are subtracted from the uncorrected wavefronts. The uncorrected mean RMS errors are 0.50~rad for the random tip/tilt and 0.52~rad for the temporal tip/tilt errors. Both networks reduce the mean tip/tilt RMS error by nearly an order of magnitude. In the random case, for the CNN, the mean RMS error is~0.07~rad, and for the TNN, the mean RMS error is~0.08~rad. In the temporal case, for the CNN, the mean RMS error is~0.07~rad, and for the TNN, the mean RMS error is~0.05~rad. The advantage of the TNN appears when the data is temporally correlated, where the TNN exploits sequence information.

Both networks are able to learn the low-order wavefront errors with high accuracy, reducing the residual tip/tilt RMS error to below 0.1~rad. However, the minor increase in estimation performance of the TNN when processing time-series data hints at its potential for wavefront sensing using temporal sequences of lantern intensity inputs.

\subsection{Von K\'arm\'an Seeing}

Assuming that there is no upstream wavefront correction, we test the performance of the CNN and TNN at estimating direct Von K\'arm\'an seeing (discussed in Section~\ref{sec:wfe}), \textcolor{black}{noting that the 6-core HMS-PL offers only low-order wavefront sensing.}

To quantify the wavefront estimation performance of the CNN and TNN when encountering Von K\'arm\'an seeing, we calculate the Strehl ratio $S$. The Strehl ratio is defined as the ratio of the peak intensity of an aberrated point-source image to the peak intensity that the same optical system would produce if it were perfectly diffraction-limited~\cite{Mahajan1983_strehl}. In practice, the Strehl ratio is often approximated using the Maréchal approximation, calculated as a function of the residual wavefront variance $\sigma^2_{\phi}$ in rad$^2$ as,
\begin{equation} \label{eq:strehl}
    S \approx \exp(-\sigma^2_{\phi}).
\end{equation}

\noindent Maréchal approximation is commonly used in AO as direct imaging of the aberrated PSFs is uncommon for on-sky wavefront correction. A Strehl ratio of~1 represents a perfect wavefront correction (flat wavefront), where increasingly aberrated wavefronts reduce the Strehl ratio to~0.

\begin{figure}[ht] 
    \centering
    \includegraphics[scale=0.55]{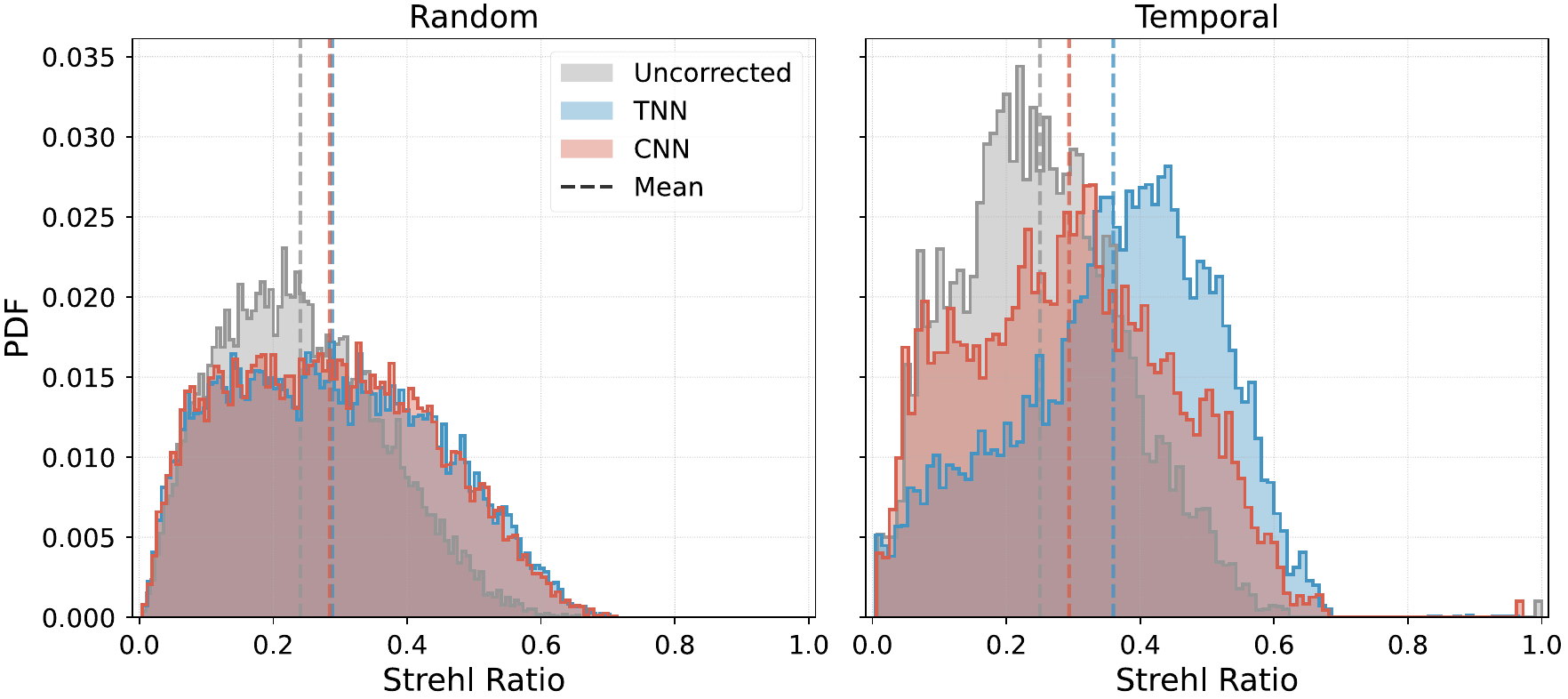}
    \caption{Strehl ratio probability density functions (PDFs) for Von K\'arm\'an seeing, both with randomized and sequential temporal data, uncorrected and after low-order wavefront estimation \textcolor{black}{used for low-order correction} with the TNN and CNN.}
    \label{fig:vk_sr}
\end{figure}

We show the probability density functions (PDFs) of the Strehl ratio for the uncorrected wavefronts, and both the random and temporal Von K\'arm\'an seeing in Figure~\ref{fig:vk_sr}. For the random wavefront instantiations, the mean Strehl ratio of the uncorrected wavefronts is $S\approx 0.24$. \textcolor{black}{When the estimated wavefronts are subtracted from the uncorrected wavefronts,} both the CNN and TNN PDFs show approximately the same estimation performance, with mean Strehl ratios of $S\approx 0.28$. For the temporal wavefront sequences, the uncorrected wavefronts have a mean Strehl ratio of $S\approx 0.25$. The CNN estimations result in a mean Strehl ratio of $S\approx 0.29$ and the TNN estimations result in a mean Strehl ratio of $S\approx 0.36$, where the TNN PDF can be seen shifted towards higher Strehl ratios. These results clearly highlight the utility of TNNs at interpreting complex temporal information.

\begin{figure}[h] 
    \centering
    \includegraphics[scale=0.55]{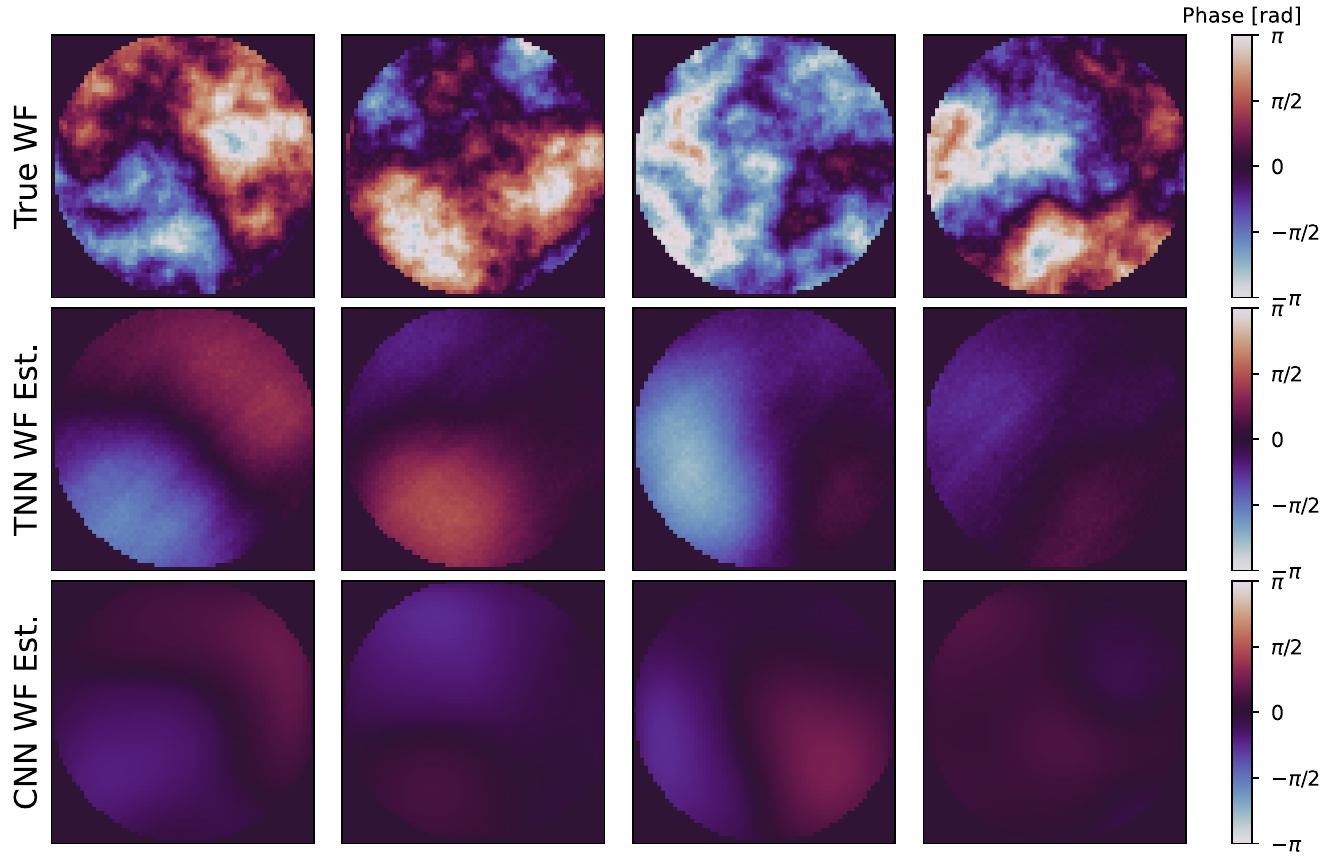}
    \caption{Example wavefronts for temporal Von K\'arm\'an seeing, uncorrected and estimated using the TNN and CNN.}
    \label{fig:wf_ex}
\end{figure}

Examples of the temporal wavefront estimations output by the CNN and TNN are given in Figure~\ref{fig:wf_ex}, where the true wavefronts are plotted along the top row, the TNN estimations are plotted along the middle row, and the CNN estimations are plotted along the bottom row.

It can be seen that the TNN estimates are able to recover the dominant low-order wavefront errors, where general large-scale gradient orientations and signs are captured; however they are much smoother than the true wavefronts, with weaker phase magnitudes. The CNN estimates are smoother than the TNN estimates, with a lower correlation with the general gradient structure of the true wavefront, and even weaker phase magnitudes. Both networks essentially act as a low-pass filter of the true wavefront.

\textcolor{black}{We emphasize that the 6-core HMS-PLs provide only low-order wavefront sensing capabilities, thus, do not represent achievable PL-based wavefront estimation performance in general (e.g., see on-sky performance of a 19-core PL wavefront sensor in~\cite{Lin2025}).}

The advantage of the TNN in handling temporal data is clearly highlighted in Figures~\ref{fig:vk_sr}~and~\ref{fig:wf_ex}, which motivates future development of TNNs for PL wavefront sensing. In particular, future work could develop TNN architectures for closed-loop, predictive adaptive optics, where the SNR problem for HMS-PLs can be explored in greater depth. \textcolor{black}{The advantage of TNNs for PL-based wavefront correction could be unlocked further when applied to PLs supporting higher-order modes, and/or with broadband spectrally dispersed intensities output from the PLs, which would provide richer information for the TNNs to learn from.}

\section{CONCLUSION}
\label{sec:concl}

We presented an analysis of two different neural network-based wavefront estimation algorithms for hybrid-mode selective photonic lantern wavefront sensing. A transformer neural network was found to outperform a convolutional neural network at learning temporal wavefront sequences, motivating their use in future closed-loop predictive wavefront correction for lantern-based wavefront sensing.



\acknowledgments 
We acknowledge support from Astralis - Australia\textquoteright s optical astronomy instrumentation Consortium - through the Australian Government\textquoteright s National Collaborative Research Infrastructure Strategy (NCRIS) Program. B.N. acknowledges support from an ARC Future Fellowship FTFT240100614. F.M acknowledges funding from the project PHOTONICS financed by the ANR program PEPR Origins (ANR-22-EXOR-0005).

\bibliography{bibliography} 
\bibliographystyle{spiebib} 

\end{document}